\documentclass[preprintnumbers,amsmath,amssymb,floatfix,11pt,prd,onecolumn, superscriptaddress,nofootinbib]{revtex4}
\usepackage{graphicx}
\usepackage{epsfig}
\usepackage{bm}
\usepackage{amsfonts}

\begin{document}

\title{Noether symmetry of F(T) cosmology with quintessence and phantom scalar fields}

 \author{\textbf{Mubasher Jamil}}\email{mjamil@camp.nust.edu.pk}
\affiliation{Center for Advanced Mathematics and Physics (CAMP),
National University of Sciences and Technology (NUST), H-12,
Islamabad, Pakistan}\affiliation{Eurasian International Center for
Theoretical Physics, Eurasian National University, Astana 010008,
Kazakhstan}

 \author{\textbf{D. Momeni}}\email{d.momeni@yahoo.com}\affiliation{Eurasian International Center
for Theoretical Physics, Eurasian National University, Astana
010008, Kazakhstan}

 \author{\textbf{R. Myrzakulov}}\email{rmyrzakulov@csufresno.edu}\affiliation{Eurasian International Center
for Theoretical Physics, Eurasian National University, Astana
010008, Kazakhstan}

\begin{abstract}
\textbf{Abstract:}
In this paper, we investigate the Noether symmetries of
$F(T)$ cosmology involving  matter and dark energy. In this model, the dark energy is represented by a
 canonical scalar field with a potential. Two special cases for dark energy are considered including
  phantom energy and quintessence. We obtain $F(T)\sim T^{3/4},$ and the scalar potential $V(\phi)\sim\phi^2$ for both models of dark energy and discuss quantum picture of this model. Some astrophysical implications are also discussed. 

\end{abstract}

\maketitle
\newpage
\section{Introduction}

In the last decade, one of the most active researches in physicist
community is investigation of the acceleration of our universe
\cite{expansion}, which has confirmed by some observation data
such as supernova type Ia \cite{Supernova}, baryon acoustic
oscillations \cite{oscillations}, weak lensing \cite{lensing} and
large scale structure \cite{LSS}. Finding the theoretical explanation of cosmic acceleration has been
one of the central problems of modern cosmology and theoretical physics \cite{smolin}.
Reviews of some recent and old attempts to resolve the issue of dark energy and related
problems can be found in \cite{reviews}.

In order to explain the current accelerated expansion without
introducing dark energy, one may use a simple generalized version of
the so-called teleparallel gravity (TG) \cite{Albert}, namely $F(T)$
theory. It is a generalization of the teleparallel gravity  by
replacing the so-called torsion scalar $T$ with $F(T)$. TG was
originally developed by Einstein in an attempt of unifying gravity
and electromagnetism. The field equations for the $F(T )$ gravity
are very different from those for $f(R)$ gravity, as they are second
order rather than fourth order. $F(T )$ gravity is not locally
Lorentz invariant and appear to harbor extra degrees of freedom not
present in general relativity \cite{li}. In fact as Li et al
\cite{li} pointed out ``there are $D - 1$ extra degrees of freedom
for $F(T)$ gravity in $D$ dimensions, and this implies that the
extra degrees of freedom correspond to one massive vector field or
one massless vector field with one scalar field.'' In another recent
investigation, Li et al \cite{Barrow} studied the cosmological
perturbation and structure formation in $F(T)$ theory and proved
that the extra degree of freedom of $F(T )$ gravity decays as one
goes to smaller scales, and consequently its effects on scales such
as galaxies and galaxies clusters are small. But on large scales,
this degree of freedom can produce large deviations from the
standard cosmological model, leading to severe constraints on the
$F(T )$ gravity models as an explanation to the cosmic acceleration.

Although teleparallel gravity is not an alternative to general
relativity (they are dynamically equivalent), but its different
formulation allows one to say: gravity is not due to curvature, but
to torsion. In other word, using tetrad fields and curvature-less
Weitzenbock connection instead of torsion-less Levi-Civita
connection in standard general relativity. We should note that one
of the main requirement of $F(T)$ gravity is that there exist a
class of spin-less connection frames where its torsion does not
vanish \cite{Barrow}. $F(T)$ theory leads to interesting
cosmological behavior and its various aspects including
thermodynamic laws, phantom crossing and inflation have been
examined in the literature \cite{F(T)}.

On the other side, we can not ignore the important role of
continuous symmetry in the mathematical physics. In particular, the
well-known Noether's symmetry theorem is a practical tool in
theoretical physics which states that any differentiable symmetry of
an action of a physical system leads to a corresponding conserved
quantity, which so called Noether charge \cite{Noether}. In the
literature, applications of the Noether symmetry in generalized
theories of gravity have been studied (see \cite{ApplNoether1} and
references therein). In additions, Noether symmetry has been used to
investigate non-flat \cite{nonflatCOS} and quantum cosmology
\cite{quantumCOS}. The symmetries of the Lagrangian lead to
conserved quantities of the theory, for instance `total energy' and
`total angular momentum'. If a generic theory does not possesses a
conserved charge, it implies that this theory has nothing to do with
physical reality. Given the fact that $F(T)$ theory has no symmetry
under Lorentz transformation, it will be interesting to know if this
theory possesses any other symmetry at all. The application of
Noether symmetries helps in selecting viable models of $F(T)$ at a
fundamental level. In particular, Wei et al calculated Noether
symmetries of $F(T)$ cosmology containing matter and found a
power-law solution $F(T)\sim\mu T^n$ \cite{ApplNoether1}. Further
they showed that if $n > 3/2$ the expansion of our universe can be
accelerated without invoking dark energy. We reconsider their model
and invoke a scalar field to produce cosmic acceleration (phantom
energy and quintessence as two separate cases) and find $F(T)\sim
T^{3/4}$. Thus we show that $F(T)\sim T^{3/4}$ coupled with a
canonical scalar field is equivalent to $F(T)\sim T^n$, $n>3/2$.

The plan of our paper is as follows:  In section-II, we write down
the Lagrangian of our model and related dynamical equations. In
section-III, we write down the Noether symmety equations and solve
them. In section-IV, we discuss $F(T)$ quantum cosmology. In
section-V, we investigate the  cosmography and numerical
cosmological implications. Finally we conclude the results in
section-VI.

\section{Our model}

Here, we try to consider Noether symmetry in $F(T)$ cosmology in the
present of matter and scalar field. The Lagrangian of our model is
\begin{equation}\label{S'}
\mathcal{S}=\int d^4x~e(\mathcal{L}_F+\mathcal{L}_m+\mathcal{L}_\phi),
\end{equation}
where $e=\text{det}(e^i_\mu)=\sqrt{-g}$, where $e_i(x^\mu)$ are
related to the metric via $g_{\mu\nu}=\eta_{ij}e_\mu^i e^j_\nu$,
where all indices run over 0,1,2,3. $\mathcal{L}_F,$ $\mathcal{L}_m$
and $\mathcal{L}_\phi$ represent the Lagrangians for gravity model,
energy-matter and the scalar field (representing dark energy)
respectively. Specifically the total action reads (in chosen units
$16\pi G=\hbar=c=1$)
\begin{equation}\label{S}
\mathcal{S}=2\pi^2\int dt~ a^3\Big[F(T)-\rho_m+\frac{1}{2}\epsilon \phi_{,\mu}\phi^{,\mu}-V(\phi)\Big].
\end{equation}
where $a$ is the scale factor while $H\equiv\dot a/a$ is the Hubble
parameter.  Here $\epsilon=+1,-1$ represent quintessence and phantom
(or ghost) dark energies respectively.  Although phantom DE is the
least desirable candidate of DE as it violates relativistic energy
conditions and leads to future time singularities, we consider it
for the sake of completeness of our model since some astrophysical
observations support it (see \cite{phantom} and references therein).
The scalar field $\phi$ has the potential energy $V(\phi)$ (to be
determined in the later sections) and $\rho_m=\rho_{m0}a^{-3}$ is
the energy density of matter with vanishing pressure and $\rho_{m0}$
is a constant energy density at some initial time.

The Friedmann-Robertson-Walker (FRW) metric representing  a
spatially flat, homogeneous and isotropic spacetime is given by
\begin{equation}\label{frw}
ds^2=-dt^2+a(t)^2(dx^2+dy^2+dz^2).
\end{equation}
In FRW cosmological background, the Lagrangian  is
\begin{eqnarray}\label{S}
\mathcal{S}&=&2\pi^2\int dt~a^3\Big[F(T)-\lambda
(T+6H^2)-\frac{\rho_{m0}}{a^3}+\frac{1}{2}\epsilon \phi_{,\mu}\phi^{,\mu}-V(\phi)\Big].
\end{eqnarray}
where $T=-6H^2$ is the torsion scalar, $\lambda$ is the  Lagrange
multiplier and can be determined by varying the Lagrangian with
respect to torsion scalar $T$, which yields $\lambda=F_T$.
Integrating by parts in (\ref{S}), the action converts to
\begin{eqnarray}\label{1111}
\mathcal{S}&=&2\pi^2\int dt \Big[a^3(F-TF_{T})-6F_T
 a\dot{a}^2-\rho_{m0}-a^3\Big(\frac{1}{2}\epsilon \dot{\phi}^2+V(\phi)\Big)\Big],
\end{eqnarray}
and then the point-like Lagrangian reads (ignoring a constant factor $2\pi^2$)
\begin{eqnarray}\label{1}
\mathcal{L}(a,\dot a,\phi,\dot\phi,T)&=& ~a^3(F-TF_{T})
-6F_T a\dot{a}^2-\rho_{m0}-a^3\Big(\frac{1}{2}\epsilon \dot{\phi}^2+V(\phi)\Big).
\end{eqnarray}
Moreover for a given dynamical system, the Euler-Lagrange equation is
\begin{equation}\label{EL}
\frac{d}{dt}\Big( \frac{\partial\mathcal{L}}{\partial\dot q_i} \Big)-\frac{\partial\mathcal{L}}{\partial q_i}=0,
\end{equation}
where $q_i=a,\phi,T$ are the generalized coordinates  of the
configuration space $\mathcal{Q}=\{a,\phi,T\}$. Using (\ref{1}) in
(\ref{EL}), we obtain the three equations of motion (corresponding
to variations of $\mathcal{L}$ with respect to $T,\phi,a$
respectively)
\begin{equation}
a^3F_{TT}( T+6H^2)=0,\label{EL1}
\end{equation}
\begin{equation}
 \epsilon(\ddot{\phi}+3H\dot{\phi})-\frac{\partial V}{\partial\phi}=0,\label{EL2}
 \end{equation}
 \begin{equation}
 4\frac{\ddot{a}}{a}(F_T+2TF_T)+4H^2(F_T-2TF_T)+F-TF_{TT}=\rho_\phi,\label{EL3}
\end{equation}
where $$\rho_\phi\equiv\frac{1}{2}\epsilon \dot{\phi}^2+V(\phi).$$
Assuming $F_{TT}\neq0$, from (\ref{EL1}) we find
\begin{equation}\label{T}
T=-6H^2,
\end{equation}
which is the torsion scalar for FRW model.  Using $\frac{\ddot
a}{a}=H^2+\dot H$ and (\ref{T}) in (\ref{EL3}), we find
\begin{equation}
48H^2F_{TT}\dot H-4F_T(3H^2+\dot H)-F=p_\phi,
\end{equation}
is the modified Raychaudhuri equation, where  $p_\phi=-\rho_\phi$
(note that $p=p_\phi,$ $p_m=0$). We also write down the Friedmann
equation for our model
\begin{equation}\label{FRE}
12H^2F_T+F=\rho_\phi+\rho_m.
\end{equation}

\section{Noether Symmetry Analysis}

The Noether symmetry approach is useful in  obtaining exact solution
to the given Lagrangian i.e. unknown functions in a given Lagrangian
can be determined up to some arbitrary constants.

The Noether symmetry generator is a vector field defined by
\begin{equation}\label{2}
\textbf{X}=\alpha\frac{\partial }{\partial a}+\beta\frac{\partial}{\partial \phi}+\eta\frac{\partial }{\partial T}+
\dot \alpha\frac{\partial }{\partial \dot a}+\dot \beta\frac{\partial }{\partial \dot\phi}+\dot \eta\frac{\partial }{\partial \dot T},
\end{equation}
where dot represents the total derivative given by
\begin{equation}
\frac{d}{dt}\equiv\dot\phi\frac{\partial}{\partial\phi}+\dot a\frac{\partial}{\partial a}+\dot T\frac{\partial}{\partial T}.
\end{equation}
The vector field $\textbf{X}$ can be thought of as a vector field on $\mathcal{T Q}=(a,\dot a,\phi,\dot\phi,T,\dot T)$ is the related tangent bundle on which $\mathcal{L}$ is defined.

A Noether symmetry $\textbf{X}$ of a Lagrangian $\mathcal{L}$ exists if the Lie derivative of $\mathcal{L}$ along the vector field $\textbf{X}$ vanishes i.e.
\begin{equation}\label{3}
L_X\mathcal{L}=\textbf{X}\mathcal{L}=\alpha\frac{\partial \mathcal{L}}{\partial a}+\beta\frac{\partial\mathcal{ L}}{\partial \phi}+\eta\frac{\partial \mathcal{L}}{\partial T}+
\dot \alpha\frac{\partial \mathcal{L}}{\partial \dot a}+\dot \beta\frac{\partial \mathcal{L}}{\partial \dot\phi}+\dot \eta\frac{\partial \mathcal{L}}{\partial \dot T}=0.
\end{equation}
By requiring the coefficients of $\dot a^2$, $\dot\phi ^2$, $\dot T^2$, $\dot a \dot\phi$, $\dot a \dot T$ and $\dot \phi \dot T$ to be zero in Eq. (\ref{3}), we find
\begin{eqnarray}
3\alpha F -3 \alpha T F_T-3\alpha V(\phi)-\eta a T F_{TT}-\beta a V'(\phi)&=&0,\label{s1}\\
\alpha F_{T}+\eta a F_{TT}+2a F_{T}\frac{\partial \alpha}{\partial a} &=&0,\label{s2}\\
\frac{3}{2}\alpha+a\frac{\partial \beta}{\partial \phi} &=&0,\label{s3}\\
12 F_T \frac{\partial \alpha}{\partial \phi}+\epsilon a^2 \frac{\partial \beta}{\partial a}&=&0,\label{s4}\\
12 a F_T\frac{\partial \alpha}{\partial T} &=&0,\label{s5}\\
\epsilon a^3 \frac{\partial \beta}{\partial T} &=&0.\label{s6}
\end{eqnarray}

By assuming $F_T\neq0$, from Eqs. (\ref{s5}) and (\ref{s6}), we conclude
\begin{equation}
\alpha=\alpha(a,\phi), \ \ \beta=\beta(a,\phi).
\end{equation}
Now we must solve the system of equations (\ref{s1})-(\ref{s4}). The non-trivial solution for this system reads as the following form (Model-I)
 \begin{eqnarray}
 F(T)&=&\frac{4}{3}c_1 T^{\frac{3}{4}}+c_3,\label{f}\\
 V(\phi)&=&c_4+c_5(c_1\phi+c_2)^2\label{v},\\
 \alpha(a,\phi)&=&-\frac{2}{3}c_1 a\label{alpha},\\
 \beta(a,\phi,T)&=&c_1\phi+c_2\label{beta},\\
 \eta(a,\phi,T)&=&\frac{8}{3}c_1 T\label{eta}.
 \end{eqnarray}

Here $c_1\ldots c_5$ are arbitrary constants. It is interesting to note that the form of potential $V$ and torsion function $F$ is the same for both models of dark energy. The quadratic potential (\ref{v}) has been used to model cosmic inflation including chaotic inflation in super-gravity models \cite{linde}.

Using (\ref{alpha})-(\ref{eta}), the Noether symmetries are
 \begin{eqnarray}
\textbf{ X}_1=-\frac{2}{3}a\frac{\partial}{\partial a}+\frac{8}{3}T\frac{\partial}{\partial T}+\phi\frac{\partial}{\partial \phi},\ \ \textbf{X}_2=\frac{\partial}{\partial T}.
  \end{eqnarray}
The symmetry $\textbf{X}_1$ represents the scaling i.e. the Lagrangian remains invariant under scaling transformation while the second symmetry $\textbf{X}_2$ shows that Lagrangian is invariant under $T$ translation.

These NS generators form a two dimensional closed algebra
 \begin{eqnarray}
 [\textbf{X}_1,\textbf{X}_2]=-\frac{8}{3}\textbf{X}_2.
\end{eqnarray}
The conjugate momenta for the variables of configuration space $\mathcal{Q}$ can be defined as
\begin{eqnarray}
p_a&=&\frac{\partial\mathcal{L}}{\partial\dot a}=-12a\dot a F_T, \label{p1}\\
p_\phi&=&\frac{\partial\mathcal{L}}{\partial\dot \phi}= \epsilon\dot a ^3\dot\phi,\label{p2}\\
p_T&=&\frac{\partial\mathcal{L}}{\partial\dot T}=0.\label{p3}
\end{eqnarray}
Notice that $p_T=0$ on account of symmetry $\textbf{X}_2$. The Noether charge of the system reads
\begin{equation}\label{Q1}
Q=\alpha p_a+\beta p_{\phi}+\eta p_T=8c_1a^2\dot a F_T+(c_1\phi+c_2)\epsilon\dot a^3\dot\phi.
\end{equation}
Using (\ref{f}) in (\ref{Q1}), we get
\begin{equation}\label{Q2}
Q=-\frac{32}{3 6^{1/4}}c_1^2(\dot a a^5)^{1/2}-a^3\epsilon \dot \phi (c_1 \phi+c_2).
\end{equation}

\begin{itemize}

\item \textit{Remark-1:} If $\epsilon=0$, the Noether charge coincides with the results reported in \cite{NSF(T)} after identifying the parameters $\mu=\frac{4}{3}c_1,$ $ n=\frac{3}{4}$.

\item \textit{Remark-2:}
One obvious solution of the system (\ref{s1})-(\ref{s6}) is
\begin{equation}
\alpha=\eta=0, \ \ \beta=\textit{constant}, \ \ V(\phi)=\textit{constant}
\end{equation}
In this case $a,T$ are cyclic coordinates and we have the following constant charge
\begin{equation}
Q=\beta p_{\phi}=-\beta \epsilon a^3 \dot \phi.
\end{equation}
This is the same as the Noether symmetry analysis of purely scalar fields  in general relativity \cite{plb2010}. For $Q=0$, $\phi=$constant, which corresponds to `cosmological constant'. But if $Q\neq 0$ then we have $\dot \phi\propto a^{-3}$ and the scalar field dilutes with the expansion of the Universe.

\item  \textit{Remark-3:} Another interesting solution of the system (\ref{s1})-(\ref{s6}) is $\epsilon=V(\phi)=0$ which in this case has been discussed in \cite{NSF(T)}. For the perfect fluids with EoS $w\neq0$, the system of Noether symmetry condition is non-integrable.

\item \textit{Remark-4:}
We find another interesting solution of a teleparallel gravity with a scalar field (Model-II)
\begin{eqnarray}
F(T)&=&-\frac{3}{16}\epsilon c_1^2T+c_4,\\
V(\phi)&=&c_5e^{-2\frac{\phi}{c_1}}(1+e^{\frac{\phi+c_2}{c_1}})^2,\\  \alpha(a,\phi)&=&\frac{2}{3}\sqrt{\frac{-c_3}{a}}\sinh \Big( \frac{\phi+c_2}{c_1} \Big),\\   \beta(a,\phi)&=&\sqrt{\frac{-c_3}{a}}\cosh \Big( \frac{\phi+c_2}{c_1} \Big),\\
\eta(a,\phi,t)&=& arbitrary.
\end{eqnarray}
The corresponding Noether charge is given by
\begin{equation}
 Q= \sqrt{\frac{-c_3}{a}} \Big[-\frac{3}{2}\epsilon c_1^2a\dot a \sinh \Big( \frac{\phi+c_2}{c_1} \Big)+\epsilon a^3\dot\phi \cosh \Big( \frac{\phi+c_2}{c_1} \Big)\Big].
 \end{equation}

\end{itemize}

\section{$F(T)$ quantum cosmology}

The Hamiltonian for a given Lagrangian reads
\begin{equation}\label{H}
\mathcal{H}=\sum_i p_i\dot q_i-\mathcal{L}.
\end{equation}
Using Eqs. (\ref{p1})-(\ref{p3}) in (\ref{H}), we get
\begin{equation}\label{H1}
\mathcal{H}=-\frac{p_a}{24aF_T}-\frac{p_\phi^2}{\epsilon a^3}-a^3(F-TF_T)+\rho_{m0}+a^3V(\phi).
\end{equation}
The Hamiltonian equations are $$ \dot q_i=\frac{\partial \mathcal{H}}{\partial p_i},\ \ \dot p_i=-\frac{\partial \mathcal{H}}{\partial q_i}.  $$
\begin{eqnarray}
\dot a&=&\{a,\mathcal{H}\},\ \ \dot p_a=\{ p_a,\mathcal{H} \},\\
\dot T&=&\{T,\mathcal{H}\},\ \ \dot p_T=\{ p_T,\mathcal{H} \}\equiv0,\\
\dot \phi &=&\{\phi,\mathcal{H}\},\ \ \dot p_\phi=\{ p_\phi,\mathcal{H} \}.
\end{eqnarray}
The Hamiltonian constraint equation $\mathcal{H}\equiv0$ is equivalent to second Friedmann equation.

The Hamilton-Jacobi equation reads
\begin{eqnarray}
-\frac{1}{24 aF_T}\frac{\partial^2S}{\partial a^2}-\frac{1}{\epsilon a^3}\frac{\partial^2S}{\partial \phi^2}
-a^3(F-TF_T)+\rho_{m0}+a^3V(\phi)=0.
\end{eqnarray}
For the quantum picture of our model, we define a wave function $\psi$ and $\frac{\partial}{\partial a_i}\rightarrow -\iota \hbar\frac{\partial}{\partial q_i}$. Then the wave equation (which is the Hamiltonian constraint) reads:
\begin{eqnarray}
-\frac{\hbar^2}{24 aF_T}\frac{\partial^2\psi}{\partial a^2}-\frac{\hbar^2}{\epsilon a^3}\frac{\partial^2\psi}{\partial \phi^2}+U(a,T,\phi)\psi=0,
\end{eqnarray}
where $U(a,T,\phi)=a^3(F-TF_T)-\rho_{m0}-a^3V(\phi).$ Solution of the above wave equation for $F(T)-$ scalar field model is not our main purpose here.

\section{Numerical results and Cosmography}

In this section, using (\ref{f}) and (\ref{v}) in (\ref{EL2}) in (\ref{EL3}), we obtain the  Euler-Lagrange equations:
\begin{eqnarray}
\epsilon \ddot{\phi}+3\epsilon H\dot{\phi}&=&2c_1c_5(c_1\phi+c_2),\label{el1}\\
4\frac{\ddot{a}}{a}(F_T+2TF_T)+4H^2(F_T-2TF_T)+F-TF_{TT}&=&\frac{1}{2}\epsilon \dot{\phi}^2+c_4+c_5(c_1\phi+c_2)^2.\label{el2}
\end{eqnarray}
We numerically solve (\ref{el1}) and (\ref{el2}) and display our results in the figures 1 to 4. From figure-1, the e-folding parameter for quintessence increases almost exponentially while for phantom, its stays flat. In figure-2, the quintessence scalar field increases while phantom field oscillates and decay with time. From figure-3, the Hubble parameter decreases from its current value to nearly unity and stays close to zero for phantom while it starts increasing for quintessence. In figure-4, the state parameter for phantom decreases to sub-negative values while for quintessence, it stays near the cosmological constant boundary.

Our model (\ref{f}) must be checked by observational parameters from cosmographical view, following the methodology presented in \cite{cosmography} we must check the following equations
\begin{eqnarray}
f(T_0)&=&6H_0^2(\Omega_{m0}-2),\label{cosmo1}\\
f'(T_0)&=&1.\label{cosmo2}
\end{eqnarray}
Explicity we have
\begin{eqnarray}
c_3&=&2H_0^2(3\Omega_{m0}-2)\label{c1},\\
c_1 &=&-6^{1/4}H_0^{1/2}\label{c4}.
\end{eqnarray}
Using (\ref{c1}) and (\ref{c4}) in (\ref{f}) it is possible to find
the values of the present value of $F(T)$ and the first derivatives
of it using the cosmographic parameters set with a given value of
$\Omega_{m0}$.

\subsection{Reduction of the equations to a single equation for scale factor $a(t)$}

In this section, using (\ref{f}) and (\ref{v}) we want to find a single equation for scale factor $a(t)$.

From (\ref{el1}) we get
\begin{equation}
\epsilon (a^{3}\dot{\phi})_t=2 c_1c_5a^3\psi.
\end{equation}
where $\psi=c_1\phi+c_2$. On the other hand, from  (\ref{Q2}) we get:
\begin{equation}
\epsilon a^{3}\dot{\phi}=-\frac{Q+\frac{32}{3 6^{1/4}}c_1^2(\dot a a^5)^{1/2}}{ \psi}=-\frac{U}{ \psi}
\end{equation}
where $U=Q+\frac{32}{3 6^{1/4}}c_1^2(\dot a a^5)^{1/2}$. Differentation of this equation gives
\begin{equation}
\epsilon(a^{3}\dot{\phi})_t=-\frac{\dot{U}\psi-U\dot{\psi}}{\psi^2}
\end{equation}
So that finally we have
\begin{equation}
2 c_1c_5a^3\psi=-\frac{\dot{U}\psi-U\dot{\psi}}{\psi^2}
\end{equation}
From (41) follows
\begin{equation}
\dot{\psi}=-\frac{c_1U}{ \epsilon a^3\psi}
\end{equation}
At least, Eqs. (43) and (44) gives
\begin{equation}
2\epsilon c_1c_5a^6\psi^4+\epsilon a^3\dot{U}\psi^2-c_1U^2=0
\end{equation}
Hence we obtain
\begin{equation}
\psi^2=\pm \frac{-\epsilon a^3\dot{U}\pm \sqrt{D}}{4\epsilon c_1c_5a^6}
\end{equation}
where
\begin{equation}
 D=a^6\dot{U}^2+8\epsilon c_1^2c_5a^6 U^2\end{equation}
For  $\psi$ we get the form
\begin{equation}
\psi=\sqrt{\frac{\sqrt{D}\mp\epsilon a^3\dot{U}}{4\epsilon c_1c_5a^6}}
\end{equation}
or
\begin{equation}
\phi=-c_2+\frac{1}{c_1}\sqrt{\frac{\sqrt{D}\mp\epsilon a^3\dot{U}}{4\epsilon c_1c_5a^6}}
\end{equation}

Now let us rewrite the equation (\ref{el2}) as
\begin{equation}
[4\dot{H}(1+2T)+0.25]c_1T^{-0.25}+c_3-c_4=
\frac{2c_1c_5U^2}{\sqrt{D}\mp\epsilon a^3\dot{U}}+\frac{ \sqrt{D}\mp \epsilon a^3\dot{U}}{4\epsilon c_1a^6}.\end{equation}
This equation is very complicated in the form of the $Y(a,\dot a, \ddot a)=0$. We will not solve this equation. From analytical view, there is no simple method for converting this equation to a simpler model. The remaining job is the numerical analysis which we done in previous section.


\section{Conclusion}

Noether symmetry analysis is a useful tool to find unknown
parameters involved in the Lagrangian. As is observed in the
literature, this approach has been used to find explicit forms of
$f(R)$ and recently $F(T)$ gravities. In this Letter, we considered
the $F(T)$ cosmology with matter and (phantom or quintessence)
scalar fields with potential function. Although phantom DE is the
least desirable candidate of DE as it violates relativistic energy
conditions and leads to future time singularities, we consider it
for the sake of completeness of our model since some astrophysical
observations support it. Although in literature, one can find
numerous forms of $F(T)$ written in an ad hoc way, while the
advantage of Noether symmetry is that it helps in calculating a
viable form of this arbitrary function. We obtained $F(T)\sim
T^{3/4},$ and the scalar potential $V(\phi)\sim\phi^2$ as a viable
candidate of dark energy. In comparison with \cite{NSF(T)} dealing
with $F(T)\sim T^n$, $n>3/2$ and matter, our results remain
consistent if $n<3/2$ and a scalar field is introduced which
produces cosmic acceleration. Further we obtained a quadratic
potential $V(\phi)\sim\phi^2$ for the scalar field. Besides DE, this
potential has applications in `chaotic' inflation model. We also
wrote the Schrodinger wave equation for our model and discussed
cosmological implications of our model. Our model can be used for the construction of the $F(T)$ Quantum Cosmology.

\subsection*{Acknowledgment}

We would like to thank S.H. Hendi for useful communications and
anonymous reviewer for enlightening comments related to this work.

\newpage
\begin{figure}
\centering
 \includegraphics[scale=0.4] {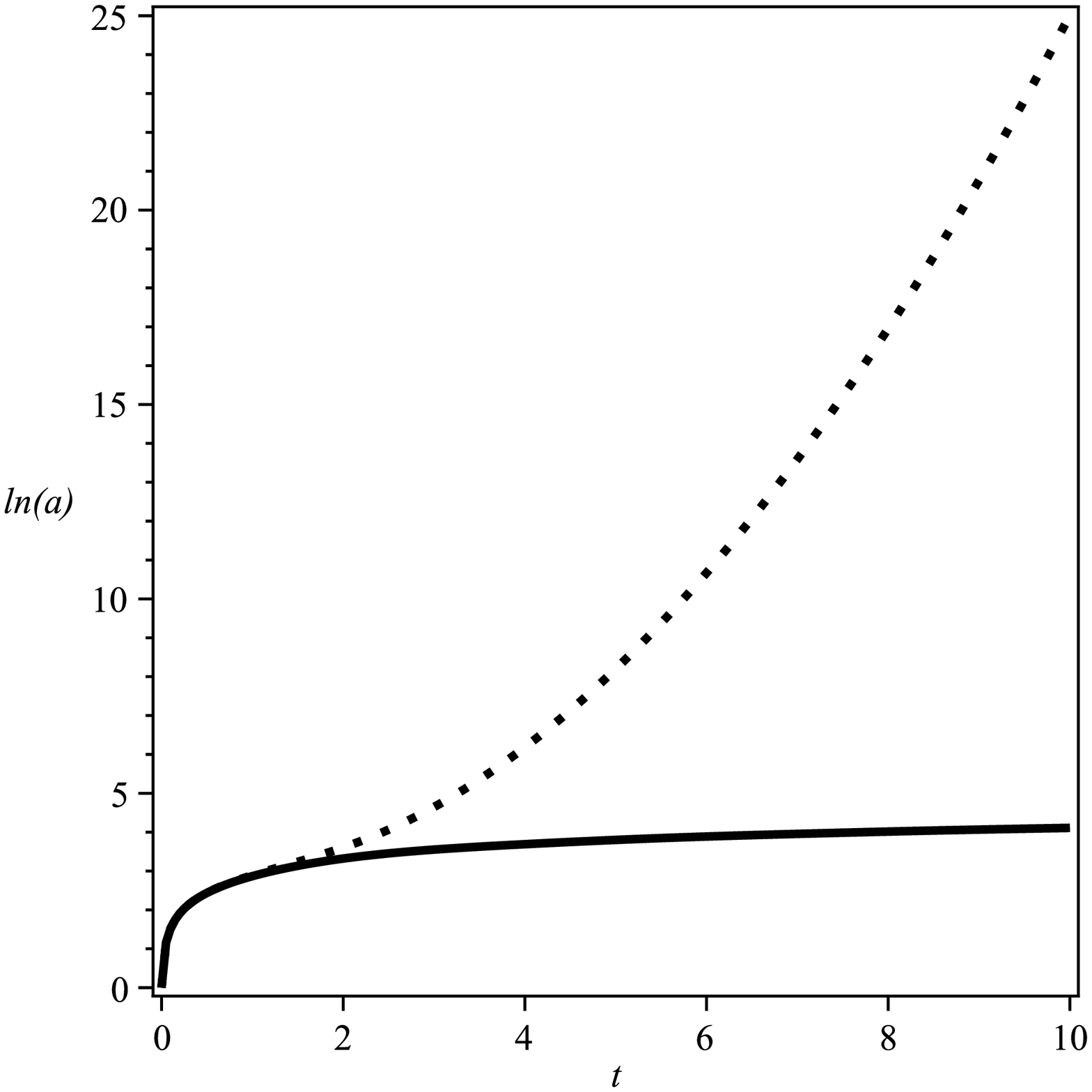}
  \caption{Graph of $ln(a)$ vs $t$. The free parameters are chosen as $c_1=\frac{3}{4},\ \ c_2=c_4=0,\ \ c_5=1$. The inital conditions are $a(0)=\phi(0)=\dot\phi(0)=1,\dot a(0)=H_0$. Here quintessence (phantom) model is shown in dot (line). }
  \label{1.eps}
\end{figure}

\begin{figure}
\centering
 \includegraphics[scale=0.4] {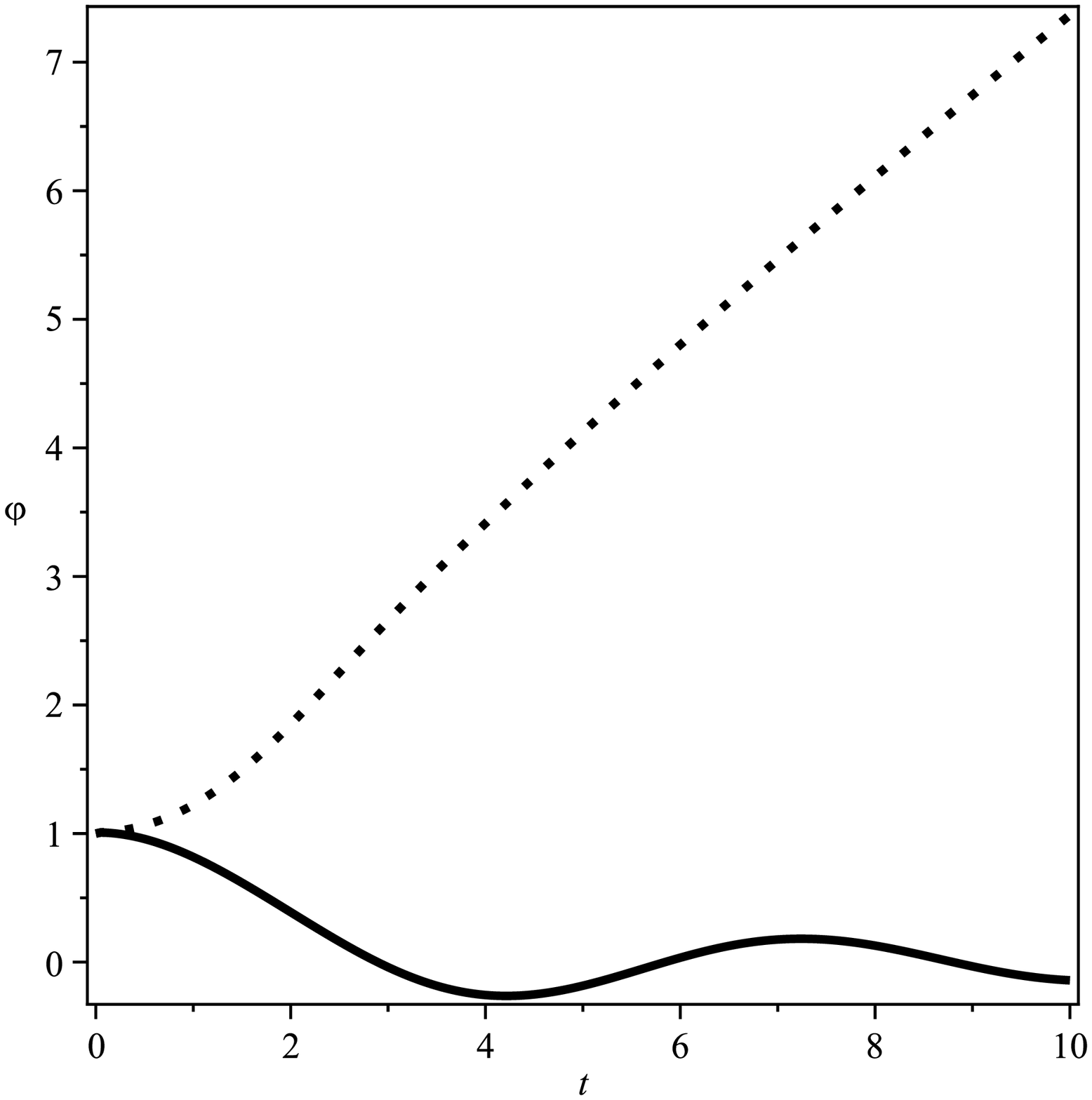}
  \caption{ Graph of $\phi(t)$ vs $t$.  The free parameters are chosen as $c_1=\frac{3}{4},\ \ c_2=c_4=0,\ \ c_5=1$. The inital conditions are $a(0)=\phi(0)=\dot\phi(0)=1,\dot a(0)=H_0$. Here quintessence (phantom) model is shown in dot (line).}
  \label{2.eps}
\end{figure}

\begin{figure}
\centering
 \includegraphics[scale=0.4] {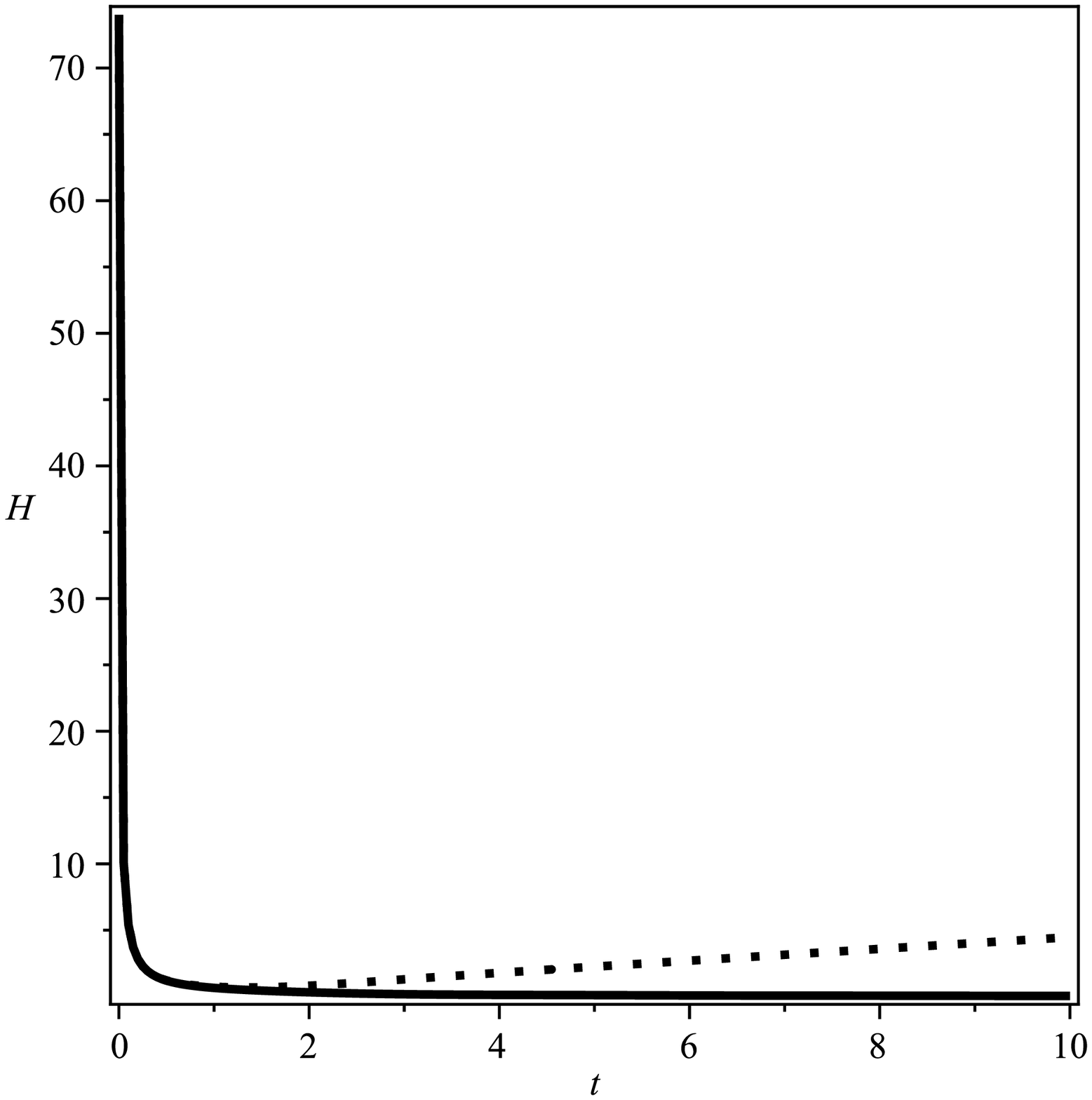}
  \caption{ Graph of $H$ vs $t$.  The free parameters are chosen as $c_1=\frac{3}{4},\ \ c_2=c_4=0,\ \ c_5=1$. The inital conditions are $a(0)=\phi(0)=\dot\phi(0)=1,\dot a(0)=H_0$. Here quintessence (phantom) model is shown in dot (line).}
  \label{3.eps}
\end{figure}

\begin{figure}
\centering
 \includegraphics[scale=0.4] {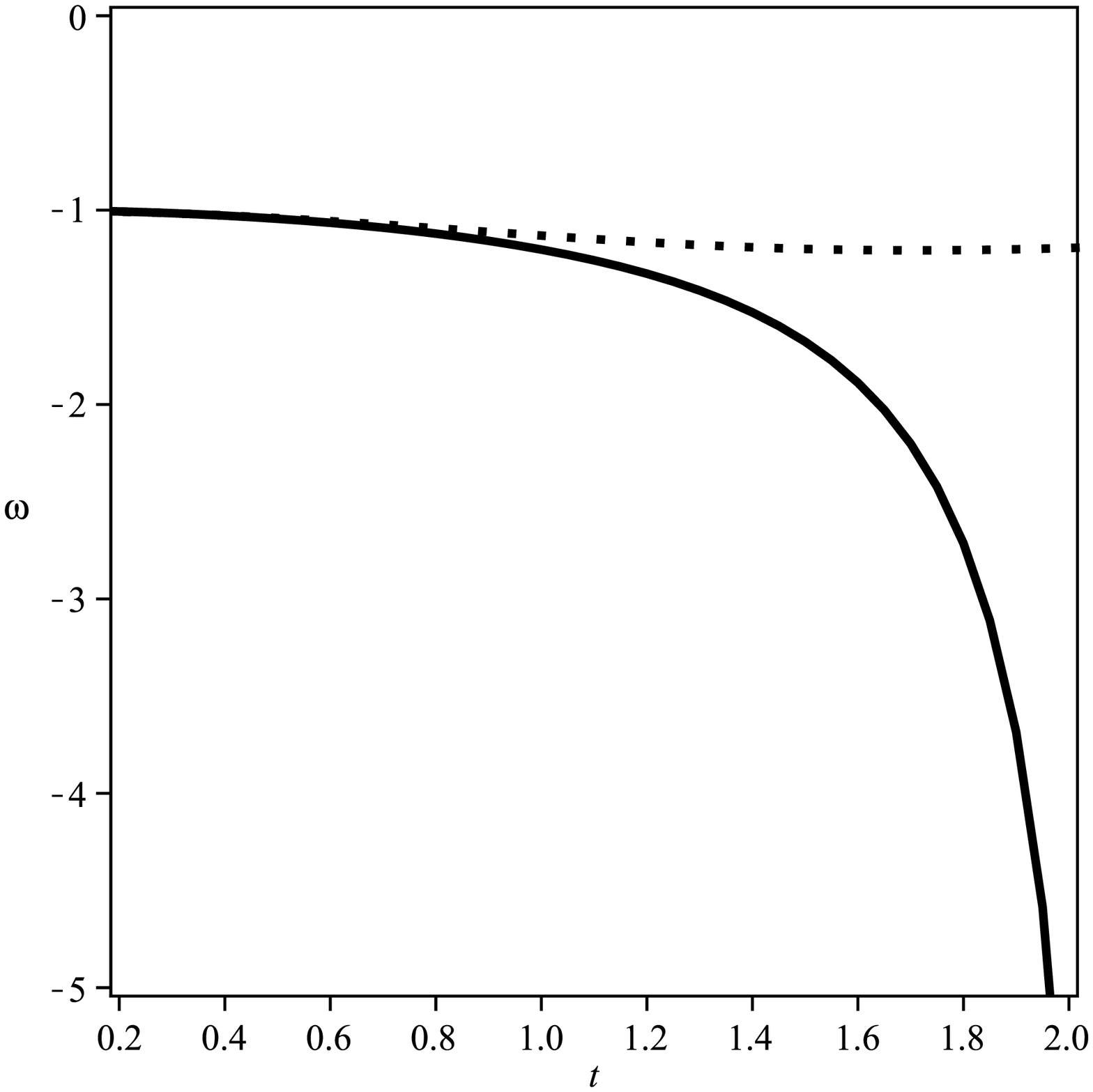}
  \caption{ Graph of $\omega$ vs $t$.  The free parameters are chosen as $c_1=\frac{3}{4},\ \ c_2=c_4=0,\ \ c_5=1$. The initial conditions are $a(0)=\phi(0)=\dot\phi(0)=1,\dot a(0)=H_0$. Here quintessence (phantom) model is shown in dot (line).}
  \label{4.eps}
\end{figure}

\end{document}